# Monolayer MoS$_2$ field effect transistor with low Schottky barrier height with ferromagnetic metal contacts

Sachin Gupta[1], F. Rortais[1], R. Ohshima[1], Y. Ando[1], T. Endo[2], Y. Miyata[2] & M. Shiraishi[1]

[1]Department of Electronic Science and Engineering, Kyoto University, Kyoto, Kyoto 615-8510, Japan

[2]Department of Physics, Tokyo Metropolitan University, Hachioji 192-0397, Tokyo, Japan

## Abstract

Two-dimensional MoS$_2$ has emerged as promising material for nanoelectronics and spintronics due to its exotic properties. However, high contact resistance at metal semiconductor MoS$_2$ interface still remains an open issue. Here, we report electronic properties of field effect transistor devices using monolayer MoS$_2$ channels and permalloy (Py) as ferromagnetic (FM) metal contacts. Monolayer MoS$_2$ channels were directly grown on SiO$_2$/Si substrate via chemical vapor deposition technique. The increase in current with back gate voltage shows the tunability of FET characteristics. The Schottky barrier height (SBH) estimated for Py/MoS$_2$ contacts is found to be +28.8 meV (zero-bias), which is the smallest value reported so-far for any direct metal (magnetic or non-magnetic)/monolayer MoS$_2$ contact. With the application of gate voltage (+10 V), SBH shows a drastic reduction down to a value of -6.8 meV. The negative SBH reveals ohmic behavior of Py/MoS$_2$ contacts. Low SBH with controlled ohmic nature of FM contacts is a primary requirement for MoS$_2$ based spintronics and therefore using directly grown MoS$_2$ channels in the present study can pave a path towards high performance devices for large scale applications.

Correspondence and requests for materials should be addressed to S.G. (email: gupta.sachin.2e@kyoto-u.ac.jp) or to M.S. (email: mshiraishi@kuee.kyoto-u.ac.jp)

# 1. Introduction

Two-dimensional (2D) materials with their layered structures have attracted much attention as next generation device materials due to their extraordinary properties such as mechanical flexibility, large surface to volume ratio, and their easy integration in heterostructure junction devices[1-3]. Graphene is well known example among these materials, which shows very rich physics resulting from its linear dispersion relation and massless Dirac Fermion[4]. Owing to remarkable properties such as very high mobility, large electrical and thermal conductivity, high Young's modulus and small spin-orbit coupling (SOC), graphene became a promising candidate for wide range of applications, including high speed electronics, sensors, energy generation and storage devices as well as spintronics[5-7]. However, semi-metallic nature (gapless band structure) of pristine graphene limits its application in semiconductor electronics as zero band-gap leads a low on/off ratio in graphene-based field effect transistors (FETs)[4,8]. In addition to this, small SOC in graphene does not allow this material to have better control on generation and electrical manipulation of spins in spintronic devices.

Unlike graphene, molybdenum disulfide ($MoS_2$), which belongs to the family of transition metal dichalcogenides (TMDs) shows semiconducting nature with a sizable band-gap[3]. The type and value of band-gap in $MoS_2$ can be changed by varying the number of layers—$MoS_2$ shows an indirect bandgap (~1.2 eV) in the bulk form (multilayers) and a direct band-gap (~1.8 eV) when reduced to monolayer[3]. In addition to non-zero band-gap, $MoS_2$ also possesses considerable SOC along with unique spin-valley coupling to manipulate the spins, which makes the material very attractive for the next generation spintronic and other technological applications. The interest in mono-layer $MoS_2$ has further increased after the demonstration of a high on/off ratio (~$10^8$) and high carrier mobility at room temperature FETs[9,10]. However, the large electrical potential drop due to high contact resistance between $MoS_2$ and metal contacts may strongly limit the performance of $MoS_2$-based devices. Previous reports show large Schottky barrier heights (SBHs) for various metal/$MoS_2$ contacts[11], which can be reduced by different approaches such as insertion of insulating layers (*h*-BN[12], MgO[13], TiO$_2$[14,15], Al$_2$O$_3$[16]) between metal and MoS$_2$, chemical doping[17,18] of MoS$_2$ and electrical gating[13,16]. To study spin injection from ferromagnetic (FM) metal and spin transport in MoS$_2$, it is very important to investigate the contact behavior between FM metal and MoS$_2$ and to suppress the SBH that hinders efficient spin injection/detection. There are only few reports in the literature, which discussed behavior of FM/ monolayer MoS$_2$ contacts and estimated SBH[12,13,16]. From our knowledge, the MoS$_2$ in previous reports is either exfoliated from the bulk single crystal of MoS$_2$ or transferred from MoS$_2$ sample grown via chemical vapor deposition (CVD) technique. However, it is notable that these methods are not suitable for mass production of electronic devices and exfoliation and/or transfer methods can induce unwanted changes in physical and electronic properties of MoS$_2$. In addition, the FET device characteristics strongly depend on the growth methods of MoS$_2$ and FM electrodes. Hence, it will be interesting to study the behavior of FM/MoS$_2$ contacts, where MoS$_2$ is directly grown on substrate by CVD technique.

In this paper, we study the device characteristics of FETs fabricated using monolayer MoS$_2$ channels, directly grown on SiO$_2$/Si substrate using salt-assisted CVD technique. 20 nm thick Ni$_{80}$Fe$_{20}$ (Py) electrodes were used as ferromagnetic contacts. The work function of Py is 4.83 eV[19]. To understand contact behavior of Py/MoS$_2$ contacts, *I-V* characteristics were studied with temperature and back gate voltages as controlling parameters. The SBH of Py/MoS$_2$ contacts was determined to be +28.8 meV at the zero-gate voltage and showed a dramatic reduction with a negative value of -6.8 meV on the application of the gate voltage of +10 V. Such contacts with low SBH and ohmic nature can play a key role in future spin-based devices because the tuning of the SBH allows circumventing the conductance mismatch problem for injecting spins in semiconductors[20].

## 2. Experimental details

The MoS$_2$ is grown on the thermally oxidized SiO$_2$/n$^+$-Si substrate with SiO$_2$ thickness of ~285 nm via salt-assisted CVD technique (please see Ref. [21] for the growth procedure). The monolayer crystal grown by this method were found to be almost free from defects/vacancies and any unintentional doping with alkali and/or halogen atoms[21]. Moreover, the sharp photoluminescence spectrum and the largest mobilities observed for these samples confirm its excellent crystal quality over exfoliated and CVD grown samples[21,22]. Figure 1(a) shows an optical microscope image of as-grown monolayer MoS$_2$, which was processed in different channels shown in Fig. 1(b). Afterwards, Py electrodes of thickness 20 nm were deposited on the top of MoS$_2$ channels, capped by Ti(3nm)/Au(50nm) as shown in Fig. 1(c). The monolayer MoS$_2$ was confirmed by the Raman spectroscopy with a laser light of wavelength of 488 nm. Only monolayer MoS$_2$ was processed for the fabrication of FET devices. Figure 1(d) shows Raman spectra for MoS$_2$ samples. The two prominent peaks in the Raman spectra appear due to an in-plane (E$_{2g}$) mode located around 382.2 cm$^{-1}$ and an out-of-plane (A$_{1g}$) mode located around 398.2 cm$^{-1}$. The difference between these two modes is found to be ~18 cm$^{-1}$, which confirms the monolayer MoS$_2$[23]. The MoS$_2$ channels [Fig. 1(b)] of length (*l*) 6-8 µm and width (*w*) 2-4 µm have been prepared by electron-beam (EB) lithography using TGMR resist (negative tone). In the first step, the MoS$_2$ channels were covered by TGMR resist and unwanted MoS$_2$ is etched by O$_2$ plasma etching for 20 seconds. In the next step, TGMR resist was lifted by wet etching using N-Methyl-2-pyrrolidone and then cleaned by acetone and isopropyl alcohol (IPA). Py(20nm)/Ti(3nm)/Au(50nm) electrodes were deposited by EB deposition technique after patterning them by EB lithography using Poly(methyl methacrylate) (PMMA) resist. The center-to-center distance between two electrodes was 0.65 µm. The electrical measurements have been performed using helium-free cryostat in the temperature range 10 -300 K. The gate voltage was applied from the backside of the Si substrate.

## 3. Results and discussion

Figure 1(e) shows the schematic for FET device and measurement configuration. Source-drain current-voltage ($I_{DS}$-$V_{DS}$) characteristics were performed by applying DC voltage and recording the current between two probes. We measured five devices (device #A1, #A2, #B1, #B2 and #C1) from three different samples (A, B and C) to check the consistency of our experimental results. In the main text, we mainly discuss FET characteristics of device #A1 while results of other devices are used as supporting information and are shown in Supplementary Material.

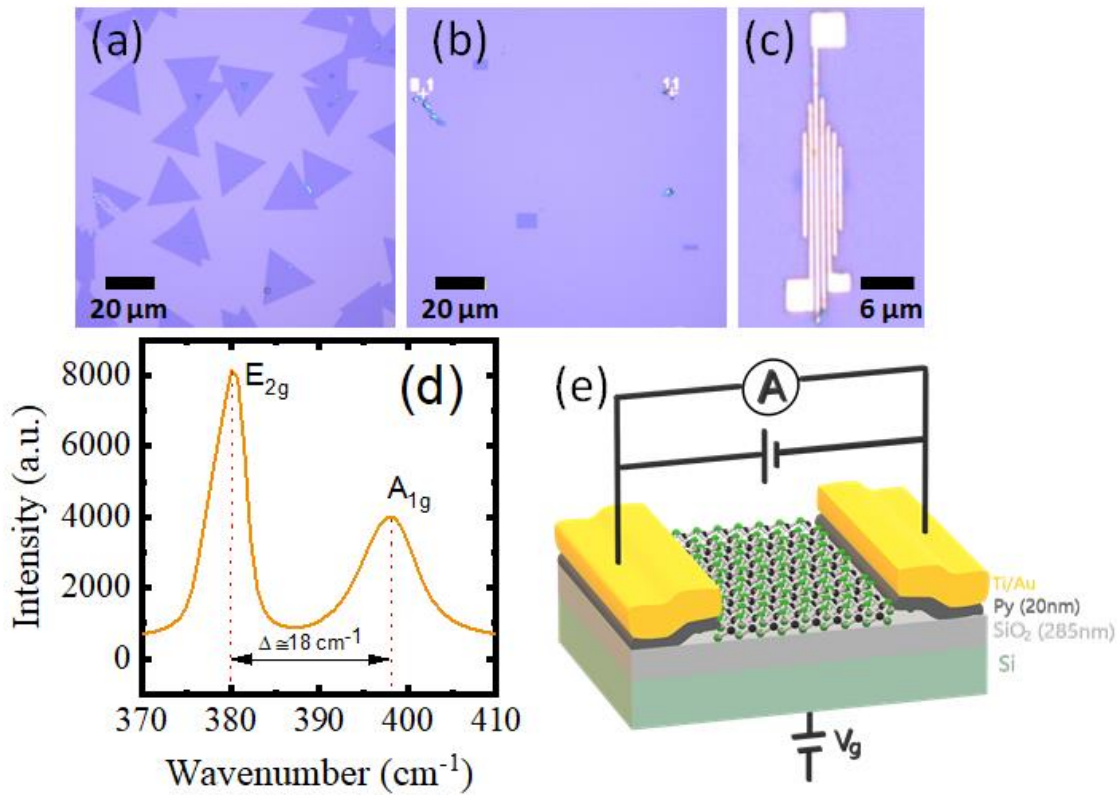

**Figure 1.** Optical microscope images of monolayer MoS$_2$—(a) As grown MoS$_2$ (triangular shaped), (b) MoS$_2$ channels of different dimensions after processing (rectangular shaped)

and (c) MoS$_2$ device with permalloy (Py) electrodes. (d) Raman spectra of monolayer MoS$_2$ performed with laser light of wavelength of 488 nm. (e) The schematic of a FET device and measurement configuration for two-probe source-drain current-voltage ($I_{DS}$-$V_{DS}$) curves.

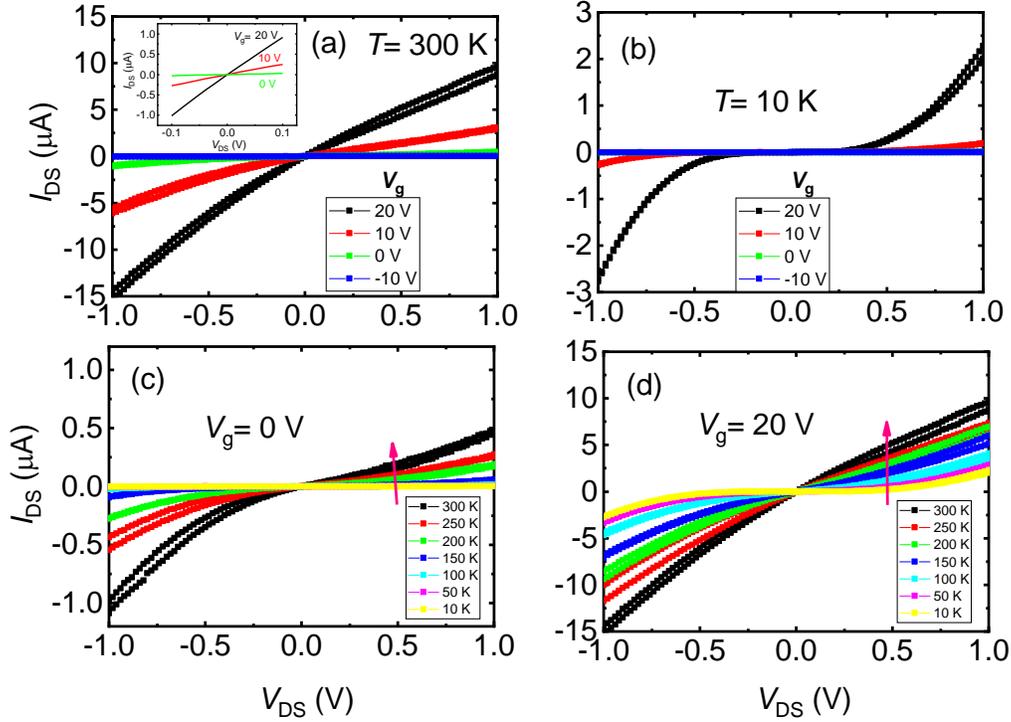

**Figure 2.** Source-drain current-voltage ($I_{DS}$-$V_{DS}$) characteristics of a MoS$_2$-based FET (a) at $T$= 300 K as a function of back gate voltage ($V_g$), (b) at $T$= 10 K as a function of $V_g$, (c) at $V_g$ = 0 V as a function of temperature, and (d) at $V_g$ = 20 V as a function of temperature. The inset in (a) shows $I_{DS}$-$V_{DS}$ characteristics under $V_{DS}$ of range $\pm 0.1\ V$. The arrow in (c) and (d) shows the direction of increase of $I_{DS}$ with increasing temperature.

To understand the electrical behavior of the Py/MoS$_2$ contacts, we carried out systematic two-probe $I_{DS}$-$V_{DS}$ curves measurements as a function of $V_g$ and temperature for device #A1, and the results are shown in Fig. 2. Figure 2(a) shows $I_{DS}$-$V_{DS}$ characteristics at 300 K as a function $V_g$. The $I_{DS}$-$V_{DS}$ curves are linear and symmetric under a small range ($\pm 0.1\ V$) of $V_{DS}$ (inset figure), however show small deviation from linearity and asymmetric behavior under high range ($\pm 1\,V$) of $V_{DS}$ (main figure). $I_{DS}$-$V_{DS}$ curves measured for device #A2 are almost linear and symmetric even under high range ($\pm 1\,V$) of $V_{DS}$ (see supplementary Fig. S1), which reflects better device quality (we note that this device was unfortunately broken during the temperature dependent measurements). It can also be noted from Fig. 2(a) that $I_{DS}$ increases with increasing $V_g$—for a fixed applied $V_{DS}$, at $V_g$= 20 V, obtained $I_{DS}$ is at least 15 times higher than that of $V_g$= 0 V. It shows the tunability of FET characteristics with the application of $V_g$ and suggests that the Schottky barrier is modified at the Py/MoS$_2$ interface, which can result in the reduction of the SBH. As temperature is lowered, $I_{DS}$-$V_{DS}$ characteristics show strong deviation from linearity and significant reduction in $I_{DS}$, as can be seen in Fig. 2(b). This suggests that the device goes in the off state at low temperatures.

Figures 2(c) and (d) show $I_{DS}$-$V_{DS}$ characteristics as a function of temperature at $V_g$ of 0 V and 20 V, respectively. The arrows in these figures show the direction of increase of $I_{DS}$ with increasing temperature. At fixed temperature and $V_{DS}$, the $I_{DS}$ is higher at $V_g$= 20 V than $V_g$= 0 V. Indeed, the Schottky barrier is lowered at 20 V, therefore more carriers can be thermally activated and overcome the barrier.

SBH can be extracted using an activation energy method, the commonly used one. The advantage of the activation energy method is that we do not need information of electrically active area under the contacts to extract the SBH[24]. The SBH for 2D materials can be extracted by employing the 2D thermionic emission equation[25,26],

$$I_{DS} = AA^*T^{\frac{3}{2}} \exp\left[-\frac{q}{k_B T}\left(\phi_B - \frac{V_{DS}}{n}\right)\right], \quad (1)$$

where $A$ is the contact surface area, $A^*$ is the effective Richardson constant, $q$ is the electronic charge, $k_B$ is the Boltzmann constant, $\phi_B$ is the Schottky barrier height and $V_{DS}$ is source-drain voltage, and $n$ is the ideality factor. From above equation, the activation energy is given by $E_A = q(\phi_B - \frac{V_{DS}}{n})$. After rearranging few terms, the equation can be written as

$$ln\left(I_{DS}/T^{\frac{3}{2}}\right) = \ln A + \ln A^* - \frac{E_A}{k_B}\left(\frac{1}{T}\right). \qquad (2)$$

From Eq. (2), it is clear that $E_A$ can be estimated from the slope of $ln\left(I_{DS}/T^{\frac{3}{2}}\right)$ vs. $1/T$, called the Arrhenius plot. Once $E_A$ is estimated, the SBH ($\phi_B$) can be extracted by simply taking the intercept of $E_A$ vs. $V_{DS}$ plot, which takes into account an effect of band bending of MoS₂ by an application of the source-drain voltage.

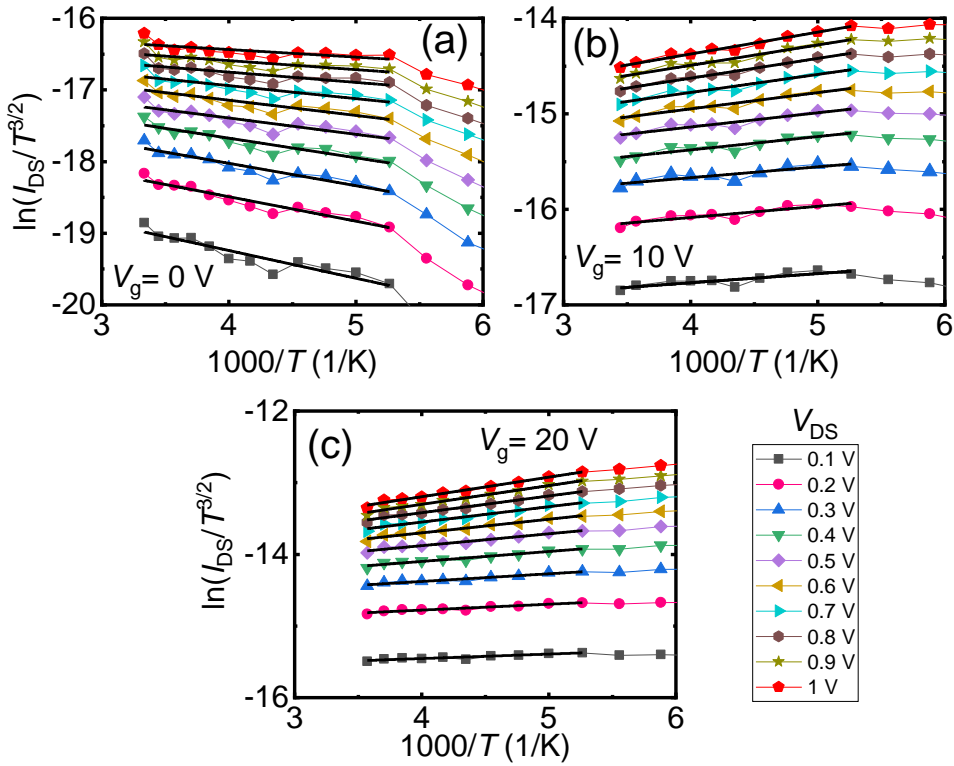

**Figure 3.** Arrhenius plots, $ln\left(I_{DS}/T^{\frac{3}{2}}\right)$ vs. $1000/T$ as a function of source-drain voltage ($V_{DS}$) for different back gate voltages, (a) $V_g$= 0 V, (b) $V_g$= 10 V, and (c) $V_g$= 20 V. The solid black line is a linear fit to Arrhenius plot to extract the slope.

$I_{DS}$-$V_{DS}$ curves recorded in high temperature range (290 -190 K) were employed to calculate the SBH. At low temperatures, the device remains in off state because the thermal energy supplied to carriers is not enough to overcome the barrier and therefore the thermionic emission theory cannot be applied successfully. Figures 3(a)-(c) show Arrhenius plots for various $V_{DS}$ at different $V_g$. To determine the slope of the plot, the experimental data were fitted by equation (2) as shown in Fig. 3 as solid black lines. It is worth to note that the slope of the Arrhenius plot for $V_g$= 0 V is negative [Fig. 3(a)] and changes its sign from negative to positive with the application of $V_g$ [Fig. 3(b and c)]. This indicates that the SBH decreases from a positive value (at $V_g$= 0 V) to negative values for $V_g$= 10 V and 20 V.

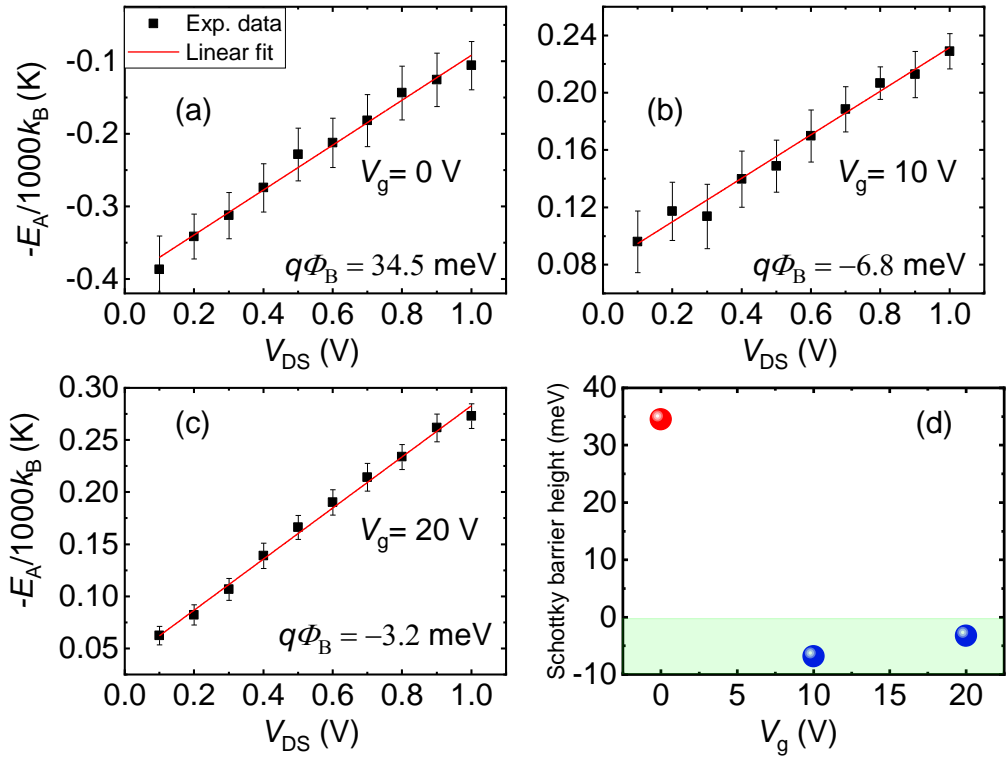

**Figure 4.** Source-drain voltage ($V_{DS}$) dependence of slopes (-$E_A$/1000$k_B$) extracted from Fig. 3 for (a) back gate voltages, $V_g = 0$ V, (b) $V_g = 10$ V and (c) $V_g = 20$ V. The solid red line is a linear fit to the experimental data to extract the intercept and therefore Schottky barrier height ($\phi_B$). (d) $V_g$ dependence of Schottky barrier height, showing positive value for $V_g = 0$ V and negative values for $V_g = 10$ and 20 V.

The slope of the Arrhenius plot as a function of $V_{DS}$ is depicted in figure 4 for different $V_g$ and fitted by linear function. The intercept from the linear fit gives the value of SBH, which is found to be +34.5 meV for $V_g = 0$ V. The SBH estimated for $V_g =$10 and 20 V from the intercept in Fig. 4(b and c) are found to be -6.8 and -3.2 meV, respectively. $I_{DS}$-$V_{DS}$ characteristics reveal that positive SBH shows depletion contact while negative SBH shows accumulation contact. Therefore, negative SBH leads to perfect ohmic contacts. To confirm the reproducibility of our results, we also estimated SBH for other FET devices fabricated on different samples. The SBHs estimated for devices #B1, #B2 and #C1 were found to be +26.0, +24.6 and 30.1 meV, respectively (see Supplementary Material), which strengthens the central claim of this study. It is quite important to estimate the SBH in TMD-based devices for future spintronic applications, because formation of the Schottky barrier strongly hinders efficient spin injection and spin detection. This is the reason we have been focusing on the SBH of the Py/MoS₂ by modulating the source-drain and gate voltages, and indeed, it is significant that we have clarified appearance of the approximately zero barrier height. Meanwhile, in a fundamental point of view, it is also important to estimate SBH at flat band gate voltage. Hence, we fabricated device #C1 and measured both the zero-bias and the flat-band SBH in the same device. As can be noted from Fig. S4 in the supplementary Material that the zero-bias and flat band SBHs are 30.1 and 21.8 meV, respectively. The observed SBH value of +28.8 meV (average value) at $V_g = 0$ V is 64% lower than the previously reported SBH in Py/MoS₂ contacts[16]. Furthermore, the ohmic regime in our devices can be realized at a smaller $V_g$ (+10 V). In the previous reports the SBH for ferromagnetic contacts such as Py/MoS₂ and Co/MoS₂ was reported to be 80.2 and 60.6 meV, where in the first case, the authors fabricated FET devices after transferring CVD grown MoS₂ to SiO₂/Si substrate via PMMA stamping method[16] and in the latter case, the FET devices were fabricated using the MoS₂ channels exfoliated from bulk MoS₂ crystals[13]. The values of SBH estimated in previous reports for various FM and non-magnetic (NM) metal contacts on monolayer MoS₂ channel are compared in Table 1. From the Table 1, it is clear that the observed SBH in our study is the smallest value reported so-far in any direct FM (non-magnetic)/monolayer MoS₂ contact. The best value reported previously for direct metal/monolayer $MoS_2$ contact, patterned by EB lithography was ~38 meV[12,27]. In fact, the SBH becomes small for transferred Ag electrodes with multilayer $MoS_2$ channels[28]. However, multilayer $MoS_2$ is not a direct band-gap semiconductor and spin-valley locking effect is somewhat suppressed[29]. Since one of the purposes of our study is exploring a potential of combination of ferromagnet and monolayer TMDs in spintronics viewpoints, realization of low SBH in Py/monolayer $MoS_2$ is crucial, although the low SBH formation to a multilayer TMD is also notable. It can be noted that $MoS_2$ channel used in previous reports were either exfoliated and/or transferred from CVD grown $MoS_2$. This implies that FET devices fabricated using $MoS_2$ directly grown on substrate via CVD technique decreases the chance of introducing distortion-induced defects during the exfoliation or the transferring methods and /or that of surface contamination, unlike exfoliated and transferred $MoS_2$ techniques. It is worth to recall that in our case $MoS_2$ is grown by salt-assisted CVD technique, which shows excellent crystal quality over exfoliated and CVD $MoS_2$ as demonstrated by the smallest SBH reported so-far in exfoliated and/or CVD $MoS_2$. High performance devices fabricated on directly grown CVD $MoS_2$ confirms the possibility of mass production of $MoS_2$-based devices. As aforementioned, circumventing the formation of the Schottky barrier is quite significant to realize efficient spin injection and detection in FM/semiconductor heterostructure, our results demonstrate the importance of an integration of directly grown $MoS_2$ channels.

**Table 1:** Comparison of Schottky barrier heights (SBHs) for direct metal (magnetic and non-magnetic) contact on monolayer MoS$_2$ channel

| Metal contact | MoS$_2$ nature | SBH (meV) | Reference |
|---|---|---|---|
| Ni$_{80}$Fe$_{20}$ (Py) | Monolayer, directly grown CVD | device #A1 = 34.5<br>device #B1 = 26.0<br>device #B2 = 24.6<br>device #C1 = 30.1<br>(flat band = 21.8)<br>**Average = 28.8** | This work |
| Ni$_{80}$Fe$_{20}$ (Py) | Monolayer, transferred CVD | 80.2 | [16] |
| Co | Monolayer, Exfoliated | 60.6 | [13] |
| Co | Monolayer, Exfoliated | ~38 | [12] |
| Ti | Monolayer, Exfoliated | 230 | [30] |
| Cr | Monolayer, Exfoliated | 130 | [30] |
| Au | Monolayer, Exfoliated | 320 | [30] |
| Pd | Monolayer, Exfoliated | 300 | [30] |

## 4. Conclusions

In conclusion, we fabricated FET devices using directly grown monolayer MoS$_2$ channels on SiO$_2$/Si substrate via salt-assisted CVD technique. The electrical properties of FET devices were studied by measuring two-probe *I-V* characteristic as a function of temperature and back gate voltage. The SBH estimated at $V_g$= 0 is found to be +28.8 meV, which is the smallest SBH reported so-far for any direct ferromagnetic as well as non-magnetic metal contact on monolayer MoS$_2$. Application of back gate voltage results in negative SBH, which indicates gate-tunable ohmic Py/MoS$_2$ contacts. Ferromagnetic contacts with the smallest SBH and controllable ohmic contacts studied in the present study can open a route for practical realization of high performance MoS$_2$ based spintronic devices.

**Data availability:** Data measured or analyzed during this study are available from the corresponding author on request.

## Acknowledgements

The authors thank Prof. T. Kimoto and K. Kanegae for their kind support in the Raman spectroscopy. This work was supported by a Grant-in-Aid for Scientific Research (S) No. 16H06330, "Semiconductor spincurrentronics", and MEXT (Innovative Area "Nano Spin Conversion Science" KAKENHI No. 26103003). Y. M acknowledges the financial support from JST CREST (grant No. JPMJCR16F3).

# Supplementary Material

# Monolayer MoS$_2$ field effect transistor with low Schottky barrier height with ferromagnetic metal contacts

Sachin Gupta,[1,*] F. Rortais,[1] R. Ohshima,[1] Y. Ando,[1] T. Endo,[2] Y. Miyata,[2] and M. Shiraishi[1]

[1]Department of Electronic Science and Engineering, Kyoto University, Kyoto, Kyoto 615-8510, Japan

[2]Department of Physics, Tokyo Metropolitan University, Hachioji 192-0397, Tokyo, Japan

*E-mail: gupta.sachin.2e@kyoto-u.ac.jp

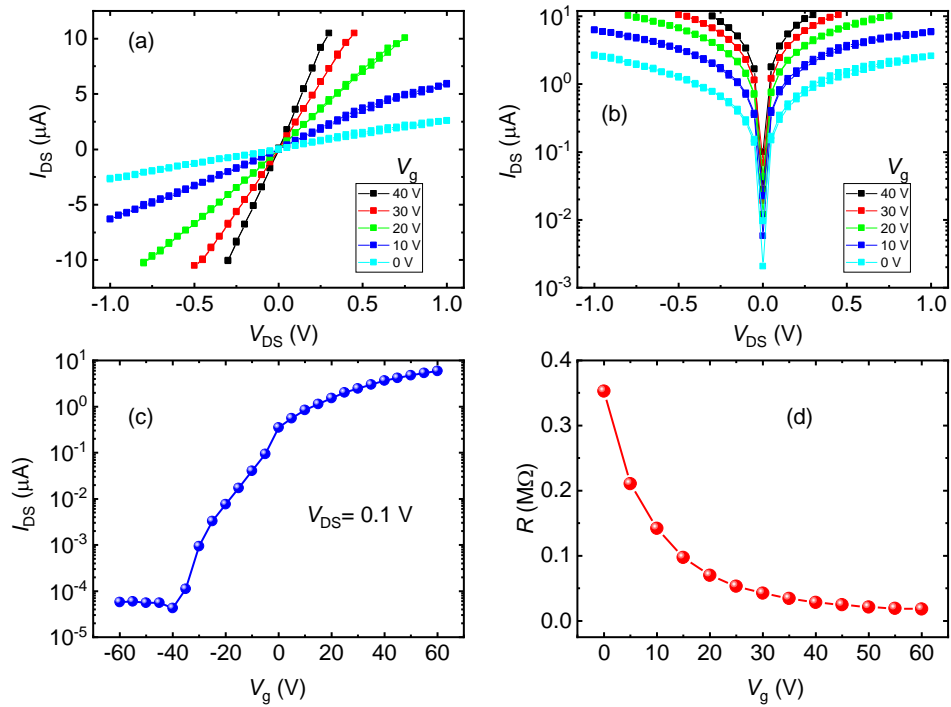

**Figure S1.** Room temperature characteristics of the FET device #A2: Source-drain current-voltage ($I_{DS}$-$V_{DS}$) characteristics of the FET plotted (a) on linear scale, (b) on semi-logarithmic scale as a function of back gate voltage ($V_g$). (c) $V_g$ dependence of $I_{DS}$ at $V_{DS} = 0.1$ V, plotted on semi- logarithmic scale. (d) $V_g$ dependence of zero bias contact resistance ($R$).

Figure S1 shows the room temperature characteristics of the device #A2. Source-drain current-voltage ($I_{DS}$-$V_{DS}$) characteristics as a function of back gate voltage ($V_g$) are shown in Figure S1(a). It can be seen that $I_{DS}$-$V_{DS}$ are linear and symmetric, suggesting ohmic contacts for this device. Figure S1(b) shows the same $I_{DS}$-$V_{DS}$ curves plotted on semi-logarithmic scale. The $V_g$ dependence of $I_{DS}$ at $V_{DS}$ = 0.1 V is plotted in Figure S1(c). $I_{DS}$ increases with increasing $V_g$ and shows an on/off ratio ~$10^5$ at $V_{DS}$ = 0.1 V. Figure S1(d) shows $V_g$ dependence of two-probe resistance (R). The $R$ shown here has two contributions: two contact resistances and the channel resistance. The $R$ decreases with $V_g$ and shows a saturation trend at $V_g$= 60V.

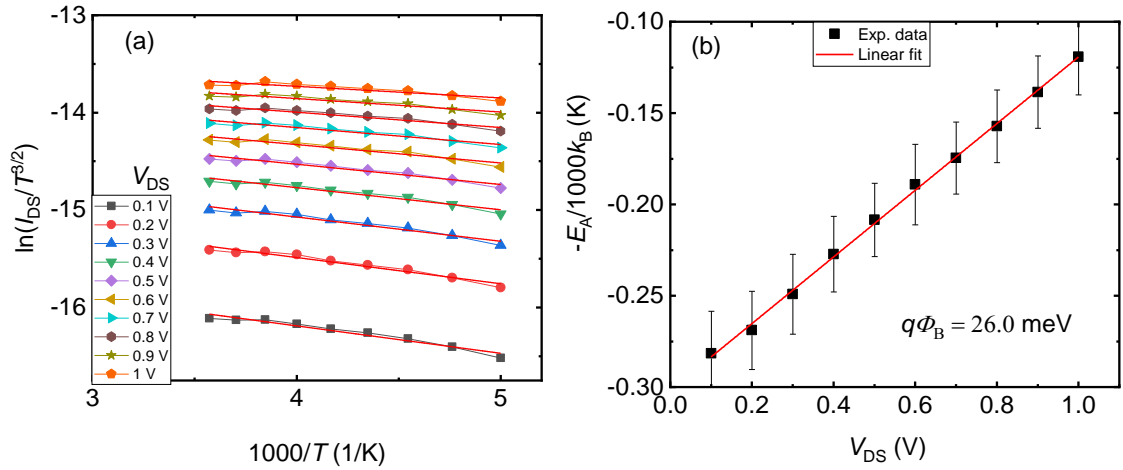

**Figure S2** Schottky barrier height calculation for the FET device #B1 at $V_g$= 0 V: (a) The Arrhenius plot, $ln\left(I_{DS}/T^{\frac{3}{2}}\right)$ vs. 1000/$T$ as a function of source-drain voltages ($V_{DS}$). (b) Source-drain voltage ($V_{DS}$) dependence of slopes (-$E_A$/1000$k_B$) extracted from (a) The solid lines are linear fits to the experimental data to extract the slope and intercept.

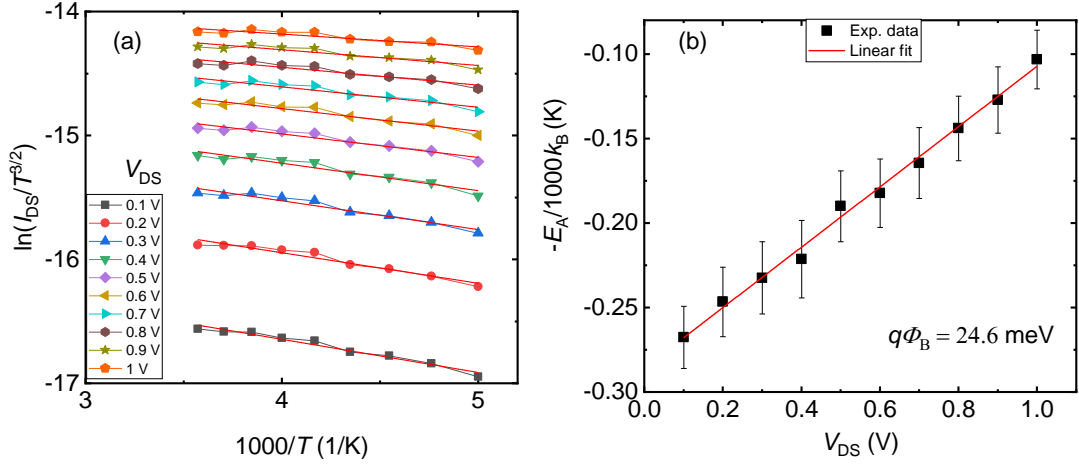

**Figure S3** Schottky barrier height calculation for the FET device #B2 at $V_g$= 0 V: (a) The Arrhenius plot, $ln\left(I_{DS}/T^{\frac{3}{2}}\right)$ vs. 1000/$T$ as a function of source-drain voltage ($V_{DS}$). (b) Source-drain voltage ($V_{DS}$) dependence of slopes (-$E_A$/1000$k_B$) extracted from (a) The solid lines are linear fits to the experimental data to extract the slope and intercept.

Figures S2(a) and S3(a) show Arrhenius plots, $ln\left(I_{DS}/T^{\frac{3}{2}}\right)$ vs. 1000/$T$ as a function of source-drain voltages ($V_{DS}$) for the devices #B1 and #B2, respectively, and figures S2(b) and S3(b) show source-drain voltage ($V_{DS}$) dependence of slopes (-$E_A$/1000$k_B$) for the devices #B1 and #B2. The procedure for estimating Schottky barrier height (SBH) is described in main text. The SBHs estimated from the devices #B1 and #B2 at $V_g$= 0 V, are found to be 26.0 and 24.6 meV, respectively.

Figure S4 shows the FET characteristics for device #C1. To calculate the SBH, we measured $I_{DS}$-$V_{DS}$ curves as a function of temperature and $V_g$. Figure S4(a) shows $I_{DS}$-$V_{DS}$ curves as a function of $V_g$ at room temperature. It can be noted that $I_{DS}$-$V_{DS}$ curves are almost linear. Figure S4(b) shows Arrhenius plots, $ln\left(I_{DS}/T^{\frac{3}{2}}\right)$ vs. 1000/$T$ as a function of source-drain voltages ($V_{DS}$). It is worth to note that with increasing $V_{DS}$, the slope changes from negative to positive, which can be clearly seen in Fig. S4(c). We plotted activation energy as a function of $V_{DS}$ and fitted the data by linear function to extract SBH. The SBH for device #C1 is estimated to be 30.1 meV at zero-bias, which is in the same range as estimated for other devices. Figure S4(d) shows $V_g$ dependence of SBH. The flat-band SBH estimated using linear fit is found to be 21.8 meV.

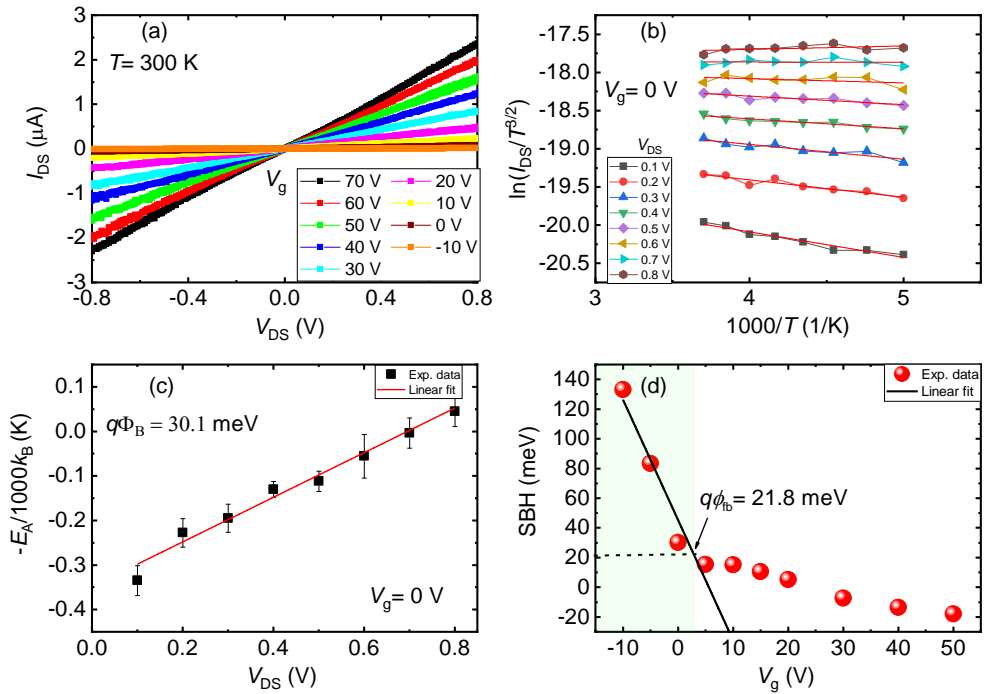

**Figure S4** Schottky barrier height (SBH) calculation for the FET device #C1 at various $V_g$: (a) Room temperature source-drain current-voltage ($I_{DS}$-$V_{DS}$) characteristics of the FET at various $V_g$. (b) The Arrhenius plot, $ln\left(I_{DS}/T^{\frac{3}{2}}\right)$ vs. 1000/$T$ as a function of source-drain voltage ($V_{DS}$). (c) Source-drain voltage ($V_{DS}$) dependence of slopes (-$E_A$/1000$k_B$) extracted from (b). The solid lines are linear fits to the experimental data to extract the slope and intercept. (d) $V_g$ dependence of SBH to extract flat band SBH.

The SBHs estimated for all four devices are almost the same, which shows consistency and reproducibility of our results and indicates that devices fabricated using MoS₂ channels directly grown on the substrate via CVD technique can result in low SBH due to less defects and/or contamination as compared to transferred or exfoliated MoS₂ as discussed in the main text.

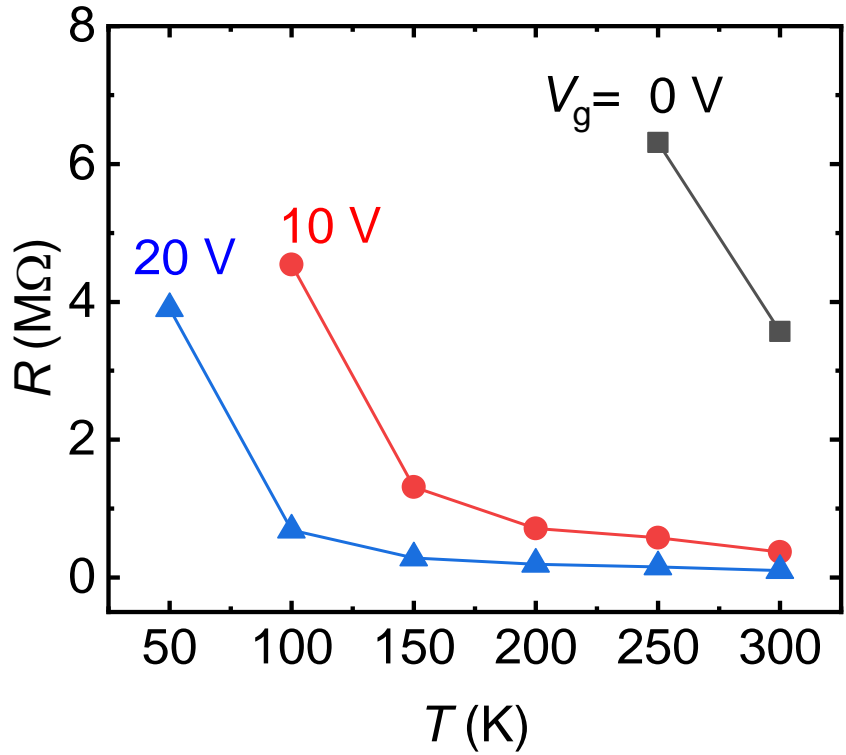

**Figure S5.** The temperature (*T*) dependence of zero-bias two-probe resistance (*R*) as a function of back gate voltage (*V*$_g$).

Figure S5 shows the temperature dependence of two-probe resistance (*R*) as a function of *V*$_g$. The *R* shown here has two contributions: two contact resistances and the channel resistance. One can note that at *V*$_g$= 0 V, the *R* is of the order of few MΩ and increases with decreasing temperature. At the low temperatures, the resistance exceeds 100 MΩ as the device goes in the off state. The *R* falls off abruptly with an application of *V*$_g$ and is almost constant (in high temperature regime), which reflects ohmic behavior of the contacts, resulting from low SBH. The *R* decreases with increasing *V*$_g$ positively, which indicates the *n*-type behavior of FET devices as expected.